\documentclass[conference]{IEEEtran}
\IEEEoverridecommandlockouts

\usepackage{cite}
\usepackage{amsmath,amssymb,amsfonts}
\usepackage{algorithmic}
\usepackage{graphicx}
\usepackage{textcomp}
\usepackage{xcolor}
\usepackage{multirow}

\def\BibTeX{{\rm B\kern-.05em{\sc i\kern-.025em b}\kern-.08em
    T\kern-.1667em\lower.7ex\hbox{E}\kern-.125emX}}
\begin{document}

\title{A Few-Shot Approach to Dysarthric Speech Intelligibility Level Classification Using Transformers}






\author{\IEEEauthorblockN{Paleti Nikhil Chowdary}
\IEEEauthorblockA{\textit{Center for Computational Engineering and Networking} \\
\textit{Amrita Vishwa Vidyapeetham}\\
Coimbatore, India \\
nikhil.28@outlook.in}
\vspace{5mm}
\IEEEauthorblockN{Vadlapudi Sai Aravind}
\IEEEauthorblockA{\textit{Center for Computational Engineering and Networking} \\
\textit{Amrita Vishwa Vidyapeetham}\\
Coimbatore, India \\
aravindvadlapudi2003@gmail.com}
\vspace{5mm}
\IEEEauthorblockN{Gorantla V N S L Vishnu Vardhan}
\IEEEauthorblockA{\textit{Center for Computational Engineering and Networking} \\
\textit{Amrita Vishwa Vidyapeetham}\\
Coimbatore, India \\
vishnuvardhangorantla0308@gmail.com}
\and
\IEEEauthorblockN{Menta Sai Akshay}
\IEEEauthorblockA{\textit{Center for Computational Engineering and Networking} \\
\textit{Amrita Vishwa Vidyapeetham}\\
Coimbatore, India \\
akshaymenta24@gmail.com}
\vspace{5mm}
\IEEEauthorblockN{Menta Sai Aashish}
\IEEEauthorblockA{\textit{Center for Computational Engineering and Networking} \\
\textit{Amrita Vishwa Vidyapeetham}\\
Coimbatore, India \\
aashishmenta249@gmail.com}
\vspace{5mm}
\IEEEauthorblockN{Jyothish Lal. G.}
\IEEEauthorblockA{\textit{Center for Computational Engineering and Networking} \\
\textit{Amrita Vishwa Vidyapeetham}\\
Coimbatore, India \\
g\_jyothishlal@cb.amrita.edu}
}

\maketitle






\begin{abstract}

Dysarthria is a speech disorder that hinders communication due to difficulties in articulating words. Detection of dysarthria is important for several reasons as it can be used to develop a treatment plan and help improve a person's quality of life and ability to communicate effectively. Much of the literature focused on improving ASR systems for dysarthric speech. The objective of the current work is to develop models that can accurately classify the presence of dysarthria and also give information about the intelligibility level using limited data by employing a few-shot approach using a transformer model. This work also aims to tackle the data leakage that is present in previous studies. Our whisper-large-v2 transformer model trained on a subset of the UASpeech dataset containing medium intelligibility level patients achieved an accuracy of 85\%, precision of 0.92, recall of 0.8 F1-score of 0.85, and specificity of 0.91. Experimental results also demonstrate that the model trained using the 'words' dataset performed better compared to the model trained on the 'letters' and 'digits' dataset. Moreover, the multiclass model achieved an accuracy of 67\%.

\end{abstract}


\section{Introduction}

Dysarthria, a neuro-motor impairment affecting speech articulation and coordination, significantly impacts an individual's ability to produce coherent and intelligible verbal communication.
In \cite{ref1} F.Rudzicz et al. claimed that dysarthria arises from congenital conditions or traumatic events that impact the neuromotor system involved in speech production. The congenital causes of dysarthria encompass conditions like brain asphyxiation during birth, which result in long-term speech impairments. On the other hand, traumatic causes of dysarthria include events such as stroke, cerebral palsy, multiple sclerosis, Parkinson's disease, myasthenia gravis, and amyotrophic lateral sclerosis (ALS). 
Individuals with dysarthria encounter difficulties related to articulation, speech rate, breath control, resonance, and overall communication \cite{ref2,padmini2022age,rajanbabu2022ensemble}. These challenges can result in diminished comprehensibility, limited expressive abilities, and obstacles in social interactions.

The field of dysarthria research has seen advancements in automatic speech recognition (ASR) systems \cite{shraddha2022child} for aiding individuals with dysarthria in communication. However, the automatic classification of dysarthria and its severity levels remain limited.
Using the Frenchay Dysarthria Assessment \cite{ref3}, doctors undertake perceptual evaluations of speech to determine the kind and severity of the disease. Subjective assessments by clinicians are costly, time-consuming, and prone to biases, raising concerns about their reliability.
This motivates the development of an impartial objective technique for evaluating dysarthric speech.

More and more researchers are employing deep learning and machine learning algorithms to develop automatic dysarthria identification in order to objectively and reliably identify individuals with the condition. 
Many researchers extract characteristics from voice signals using various feature extraction techniques \cite{lal2018epoch}. For example, Stephanie et al. \cite{ref4} used Teager Energy Operator (TEO) and the glottal waveform features. Chitralekha et al. \cite{ref5} utilized audio descriptors or features that are often used to determine the timbre of musical instruments. Dong et al. \cite{ref6} and Amlu et al. \cite{ref7} used MFCC-based features. N.P. Narendra et al. \cite{ref8} used Two sets of glottal features and acoustic features. Then, deep learning and machine learning techniques, including convolutional neural networks (CNNs), artificial neural networks (ANNs), CNN-LSTM (long short-term memory), CNN-GRU (Gated Recurrent Unit), SVM, and other models, are used to detect dysarthria.

This research aims to develop an automatic tool that leverages vocal acoustics to detect the presence of dysarthria and accurately determine its severity level. Additionally, we investigate the efficacy of different speech tasks, such as words, letters, and digits, in training the detection model. Furthermore, we explore the feasibility of employing transformer models in pathology detection, specifically dysarthria, utilizing few-shot transfer learning techniques \cite{keshav2023multimodal}.
The training process utilizes a portion of the UASpeech Dataset \cite{ref9}, while the remaining dataset is reserved for testing purposes. Log Mel spectrogram features are extracted from the audio files and are employed for training the Whisper Model \cite{ref10} which is a large language model, trained on 680,000 hours of multilingual audio data procured from the internet.
The whisper model family comprises of five different models with varying model sizes. The large variant was considered in this research which has 1550 million parameters.
Considering the computational complexity involved in training models of enormous size, various efficient training approaches were considered and LORA \cite{ref11} was used to make the training process efficient and cost-effective.  

The rest of the paper is organized as follows. Section 2 describes related works while Section 3 gives a detailed description of the methodology used. Section 4 presents the results and discussion and we conclude in Section 5.

\section{Related Works}

\begin{table*}[h!]
  \centering
  \caption{Literature Review}
  \label{tab:lit_rev}
  \begin{tabular}{|c|p{3cm}|p{2cm}|p{1cm}|p{3cm}|p{3cm}|}
    \hline
    \textbf{Reference No} & \textbf{Title} & \textbf{Author and Year} & \textbf{Dataset} & \textbf{Methodology} & \textbf{Results}\\
    \hline
    \cite{ref4} & Cross-Database Models for the Classification of Dysarthria Presence & Stephanie Gillespie, Yash-Yee Logan, Elliot Moore, Jacqueline Laures-Gore, Scott Russell, Rupal Patel (2017) & UA SPEECH, AMSDC & Prosodic, spectral, Teager Energy Operators (TEO), and glottal features of various types were recovered. To further minimize the size of the feature subsets, 10-fold cross-validation sequential forward feature selection (SFFS) was utilized. SVM is employed for categorization.& Using prosodic characteristics, the UA-Speech dataset maximizes word and participant-level accuracy at 75.3\% and 92.9\%, respectively.\\
    \hline

    \cite{ref5} & Automatic Assessment of Dysarthria Severity Level Using Audio Descriptors & Chitralekha Bhat, Bhavik Vachhani, Sunil Kumar Kopparapu (2017) & UA Speech, TORGO Database & 10 different sets of features called audio descriptors including STFT and Harmonic based features can be extracted using Multi-taper spectral estimation. An Artificial Neural Network (ANN) was used as the classifier for dysarthria severity classification & Accuracy of 96.44\% for UA speech corpus and 98.7\% for TORGO database.\\
    \hline

    \cite{ref6} & Dysarthria Speech Detection Using Convolutional Neural Networks with Gated Recurrent Unit & Dong-Her Shih, Ching-Hsien Liao, Ting-Wei Wu, Xiao-Yin Xu, Ming-Hung Shih (2022) & UA SPEECH & The model combines Convolutional Neural Networks (CNNs) and Gated Recurrent Units (GRUs). These methods allow for faster detection of dysarthria patients & The findings demonstrate that the suggested CNN-GRU model outperforms the CNN, LSTM, CNN-LSTM, and models developed by other researchers, with the greatest accuracy of 98.38\%.\\
    \hline

    \cite{ref7} & Automated Dysarthria Severity Classification:A Study on Acoustic Features and Deep Learning Techniques & Amlu Anna Joshy, Rajeev Rajan (2022) & UA SPEECH, TORGO Database & The MFCC's are computed framewise during the feature extraction stage. In the classification phase, four deep learning techniques DNN, CNN, LSTM and GRU are used.& DNN gave 93.97\% accuracy for speaker-dependent scenario and 49.22\% for speaker-independent scenario\\
    \hline

    \cite{ref8} & Dysarthric speech classification from coded telephone speech using glottal features & N P Narendra, Paavo Alku (2019) & UA SPEECH, TORGO Database & Support vector machine classifiers are trained using the suggested technique using both acoustic and glottal characteristics that were collected from coded speech utterances and their corresponding healthy/dysarthric labels. Individual and combined glottal and auditory characteristics are used to train different classifiers. & An Accuracy of 95.4\% and 96.38\% is obtained for both datasets in different instances with different feature sets\\
    \hline
    \end{tabular}
\end{table*}

There have been numerous techniques and models developed to predict the presence of dysarthria. Some of the approaches are discussed in this section and Table \ref{tab:lit_rev} presents the overview of the literature review.

In \cite{ref4}, Stephanie et al. employed a cross-database training strategy in their study to distinguish speech samples with and without dysarthria. Specifically, they trained their model on the UA-Speech database and evaluated its performance on the AMSDC database. To mitigate the issue of repeated speech samples from the same individual, one channel per participant was randomly selected for analysis.
The current analysis contains elements based on the Teager Energy Operator (TEO) and the glottal waveform in addition to conventional spectral and prosodic aspects. Baseline findings employing prosodic features on the UA-Speech dataset to optimize word and participant-level accuracy at 75.3\% and 92.9\%. However, the UA-Speech cross-training evaluated on the AMSDC maximizes word- and participant-level accuracy at 71.3\% and 90\%, respectively, based on TEO features.

In \cite{ref5}, Chitralekha et al. adopted audio descriptors or features commonly employed to characterize the timbre of musical instruments and adapted them for the purpose of their study. They utilized a dataset consisting of  dysarthric utterances, including utterances associated with 10 digits and 19 computer commands, collected from all patients. Features based on multi-tapered spectral estimates were calculated and employed for classification.  With the use of the TORGO database and the Universal Access dysarthric speech corpus, an Artificial Neural Network (ANN) was trained to categorize speech into different severity levels. For the UA speech corpus and the TORGO database, average classification accuracy was 96.44\% and 98.7\%, respectively.

In \cite{ref6}, Dong et al. used features based on MFCC Coefficients They utilized a dataset consisting of  dysarthric utterances, including utterances associated with numbers 1 to 10, the 26 letters, collected from all patients.  ,  and they used Convolutional Neural Networks (CNNs) and Gated Recurrent Units (GRUs) to enable faster dysarthria detection. Their experimental results demonstrate that the CNN-GRU model achieves an accuracy of 98.38\%, surpassing the performance of other models like CNN, LSTM, and CNN-LSTM.

In \cite{ref7} Amlu et al. employ the deep neural network (DNN), the convolutional neural network (CNN), and the  gated recurrent units(GRU) Long short term memory (LSTM) to classify the severity of dysarthric speech. Mel frequency cepstral coefficients (MFCCs) and their derivatives are the characteristics used in this investigation. For the UA-Speech database, they used 4,500 test files and 6,975 training files. Using the UA-Speech corpus and the TORGO database, The findings show that  DNN gave 93.97\% accuracy for speaker-dependent scenarios and 49.22\% for speaker-independent scenarios.

In \cite{ref8} N.P. Narendra et al. suggested a unique technique for classifying dysarthric speech from coded telephone voice using glottal characteristics. Each speaker's spoken utterances were utilized. calculated using a glottal inverse filtering technique based on deep neural networks. The openSMILE toolbox is used to integrate glottal information (time- and frequency-domain parameters and PCA-based parameters) with acoustic characteristics. Glottal and auditory characteristics are used to train both separate and mixed support vector machine classifiers. Studies using the TORGO and UA-Speech databases show that the glottal factors produced a classification accuracy range of 63–77\%.In \cite{ref14} Amlu Anna Joshy et al. also classified dysarthria using multi-head attention.

It was clear from the above literature that many methods had data leakage as audio files from the same patient were split across train and test sets. And there has not been much research conducted on few-shot learning techniques for pathology classification, which is important because the amount of audio data for pathology tasks is limited. The novelty of this work lies in exploring the effectiveness of the few-shot learning approach using transformer models like whisper-large-v2 for dysarthria detection and comparing which dataset task (Words or letters and digits) performs the best.

\section{Methodology}

\subsection{Dataset}

The goal of the UA-Speech database \cite{ref9} is to encourage the creation of user interfaces for talkers who have spastic dysarthria and severe neuromotor diseases. It consists of isolated-word recordings made using a 7-channel microphone array mounted on top of a computer display from 15 dysarthric speakers and 13 control speakers. Age, Gender, and Speech intelligibility of speakers list the dataset's varied dysarthric speakers' levels of intelligibility. This is represented in the table \ref{tab: intelligibility of speakers}.

\begin{table}[htbp]
  \centering
  \caption{Speaker-wise severity distribution for UASPEECH Database}
  \label{tab: intelligibility of speakers}
  \begin{tabular}{|c|c|c|c|}
    \hline
    \textbf{Speaker} & \textbf{Gender} & \textbf{Age} & \textbf{Speech Intelligibility (\%)}  \\
    \hline
    M09 & Male & \ 18 & High (86\%) \\
    M14 & Male & \ 44 & High (90\%) \\
    M10 & Male & \ 21 & High (93\%) \\
    M08 & Male & \ 28 & High (95\%) \\
    F05 & Female & \ 22 & High (95\%) \\
    M05 & Male & \ 21 & Mid (58\%) \\
    M11 & Male & \ 48 & Mid (62\%) \\
    F04 & Female & \ 18 & Mid (62\%) \\
    M07 & Male & \ 58 & Low (28\%) \\
    F02 & Female & \ 30 & Low (29\%) \\
    M16 & Male & \ 40 & Low (43\%) \\
    M04 & Male & \ $>18$ & Very Low (2\%) \\
    F03 & Female & \ 51 & Very Low (6\%) \\
    M12 & Male & \ 19 & Very Low (7\%) \\
    M01 & Male & \ $>18$ & Very Low (17\%) \\
\hline        
  \end{tabular}
\end{table} 

Each patient has a total of 765 files which is comprised of 10 Digits of 3 repetitions, 26 Letters of 3 repetitions, 19 Computer Commands of 3 repetitions, 100 Common Words of 3 repetitions, and 300 Uncommon Words of 1 repetition. 

For various experiments conducted in this study, various subsets of the dataset are considered. First, a dataset is prepared for the purpose of building binary classification models. This dataset is constructed by exclusively using only the common words and uncommon words of the speakers. A single repetition of common words (100 words) and all uncommon words (300 words) are combined together. In order to avoid data leakage, files from two control patients and files from two pathology patients are used for training, and files from all other patients are used for testing. The training set contained a total of 1600 audio files (800 control and 800 pathology) and the test set contained a total of 9,600 files (4400 control and 5200 pathology). 
various experiments are conducted by considering pathology patients with various intelligibility levels. A detailed description of this data is presented in table \ref{tab: Data Split for Binary Class}.

\begin{table}[ht]
    \centering
    \caption{Patient distribution of Binary Class Dataset}
    \label{tab: Data Split for Binary Class}
    \begin{tabular}{|p{2cm}|p{1cm}|p{1.05cm}|p{1cm}|p{1.05cm}|}
        \hline
        \textbf{Experiment} & \multicolumn{2}{c|}{\textbf{Train}} & \multicolumn{2}{c|}{\textbf{Test}}\\
        \cline{2-3}
        \cline{4-5}
        \textbf{Name} & \textbf{Control Patients} & \textbf{Pathology Patients} & \textbf{Control} & \textbf{Pathology}\\
        \hline
        \multirow{3}{*}{High Model} & 800 & 800 & 4400 & 5200 \\
                                   & (CM01 & (F05 &   &  \\
                                   & ,CF02) & ,M14) &   &  \\ 
                                   \hline
        \multirow{3}{*}{Medium Model} & 800 & 800 & 4400 & 5200\\
                                      & (CM01 & (F04 &   &  \\
                                      & ,CF02) & ,M11) & & \\ 
                                      \hline
        \multirow{3}{*}{Low Model} & 800 & 800 & 4400 & 5200\\
                                   & (CM01 & (F02 &   &  \\
                                   & ,CF02) & ,M16) & & \\
                                   \hline
        \multirow{3}{*}{Very Low Model} & 800 & 800 & 4400 & 5200\\
                                        & (CM01 & (F03 &   &  \\
                                        & ,CF02) & ,M12) & & \\ 
                                        \hline
    \end{tabular}
\end{table}


For the purpose of determining which dataset task gives better accuracy for multiclass models. A new dataset is created using the letters and numbers audio files. Each patient had 36 files ( 26 letters + 10 numbers) and again to avoid data leakage only two patients were considered in the training set of each class and all other patients were considered in the test set. The training set contained a total of 360 audio files (72 control and 288 pathology) and the test set contained a total of 648 audio files (396 control and 252 pathology). The detailed description of the multiclass dataset is presented in table \ref{tab: Multiclass Digits and Letters}.

                                
\begin{table}[ht]
    \centering
    \caption{Patient distribution of Multiclass Dataset using Common and Uncommon words}
    \label{tab: Multiclass}
    \begin{tabular}{|c|c|c|c|}
        \hline
        \textbf{Data Split} & \textbf{Class} & \textbf{Patients} & \textbf{Patient Files}\\
        \hline
        \multirow{5}{*}{Train} & Control & 2(CM01,CF02) & 800 \\
        \cline{2-4}
                                & High & 2(M14,F05) & 800\\
        \cline{2-4}
                                & Medium & 2(M11,F04) & 800\\
        \cline{2-4}
                                & Low & 2(M16,F02) & 800\\
        \cline{2-4}
                                & Very-Low & 2(M12,F03) & 800\\
        \hline
        \multirow{5}{*}{Test} & Control & 11 & 396 \\
        \cline{2-4}
                                & High & 3 & 1200\\
        \cline{2-4}
                                & Medium & 1 & 400\\
        \cline{2-4}
                                & Low & 1 & 400\\
        \cline{2-4}
                                & Very-Low & 2 & 800\\
        \hline
    \end{tabular}
\end{table}

                                

\begin{table}[ht]
    \centering
    \caption{Patient distribution of Multiclass Dataset using Digits and Letters}
    \label{tab: Multiclass Digits and Letters}
    \begin{tabular}{|c|c|c|c|}
        \hline
        \textbf{Data Split} & \textbf{Class} & \textbf{Patients} & \textbf{Patient Files}\\
        \hline
        \multirow{5}{*}{Train} & Control & 2(CM01,CF02) & 72 \\
        \cline{2-4}
                                & High & 2(M14,F05) & 72\\
        \cline{2-4}
                                & Medium & 2(M11,F04) & 72\\
        \cline{2-4}
                                & Low & 2(M16,F02) & 72\\
        \cline{2-4}
                                & Very-Low & 2(M12,F03) & 72\\
        \hline
        \multirow{5}{*}{Test} & Control & 11 & 4400 \\
        \cline{2-4}
                                & High & 3 & 108\\
        \cline{2-4}
                                & Medium & 1 & 36\\
        \cline{2-4}
                                & Low & 1 & 36\\
        \cline{2-4}
                                & Very-Low & 2 & 72\\
        \hline
    \end{tabular}
\end{table}

All of the input audio samples are resampled to 16,000 Hz for data preprocessing, and a representation of an 80-channel log magnitude Mel spectrogram is produced on 25-millisecond windows with a stride of 10 milliseconds. The whisper models are trained using this representation of the preprocessed data.

\subsection{Whisper Model}

\begin{figure}[htbp]
    \centering
    \includegraphics[width=0.7\linewidth]{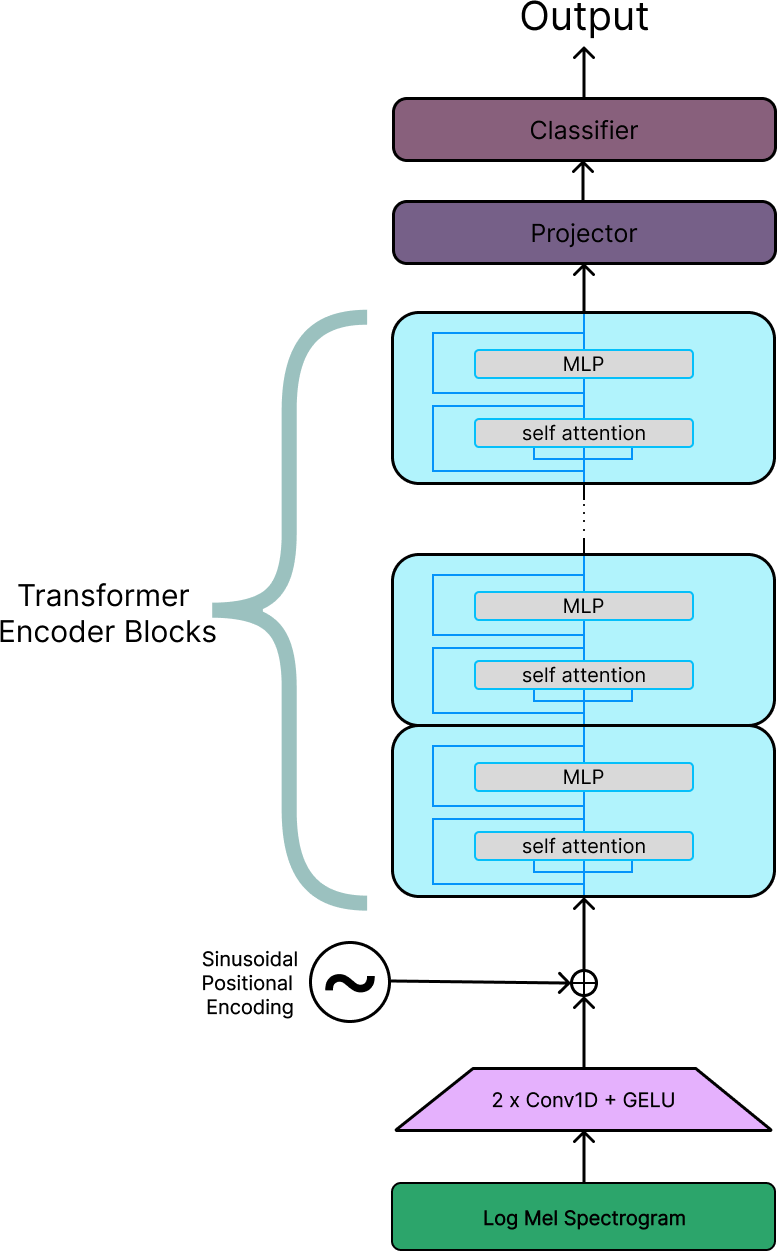}
    \caption{Modified Whisper Architecture}
    \label{fig: Working of whisper}
\end{figure}

Whisper \cite{ref10} is an Automatic Speech Recognition (ASR) system developed by OpenAI. It was trained using 680,000 hours of supervised, multilingual, and multitasking web data. The details about various architectural parameters of the whisper family models is presented in table \ref{tab: Whisper family model}. Since whisper was trained with the intention of achieving high-quality results in a zero-shot setting, this makes it very powerful and able to handle a wide range of tasks, including speech recognition. Whisper is an encoder-decoder-based architecture but since the task at hand requires only the encoder part of the model, we extracted it and added a classification head as seen in Fig. \ref{fig: Working of whisper}. Using the classification head, Given the log mel spectrogram of a speech uttered by a subject, the model will predict the probability that the subject has dysarthria and also the level of it in case of multiclass classification. 

\begin{table}[htbp]
\centering
\caption{Whisper Model Family}
\label{tab: Whisper family model}
\begin{tabular}{|c|c|c|c|c|}
\hline
Model & Layers & Width & Heads & Parameters \\
\hline
Large & 910 & 1280 & 64 & 1550M \\
Modified Large(only encoder) & 361 & 1280 & 32 & 642M \\
\hline
\end{tabular}
\end{table}

\subsection{PEFT and LoRA}

Training large language models typically requires huge clusters of GPUs and vast amounts of data. In order to make training accessible for everyone, various techniques are explored and presented. parameter-efficient fine-tuning (PEFT) \cite{ref12} selectively updates a subset of the model's parameters, specifically targeting the most influential ones for the new task. This approach significantly reduces the computational resources required for fine-tuning, resulting in improved efficiency without compromising performance. By focusing on updating only the essential parameters, we ensured effective training while minimizing unnecessary computations.

Among various methods included in PEFT, LoRA (LOW-RANK ADAPTATION OF LARGE LANGUAGE MODELS) \cite{ref11}, developed by Microsoft, is by far the most popular method. It is a technique used to fine-tune large language models (LLMs) by freezing most of the parameters and updating only a small subset specific to the task. It achieves parameter reduction by employing singular value decomposition (SVD), which decomposes a matrix into three matrices. By retaining a reduced number of singular values and their corresponding vectors, the LLM can be efficiently fine-tuned while maintaining performance.

We utilized INT8 tuning along with PEFT, LoRA and bitsandbytes \cite{ref13}.  This approach optimized memory usage and improved training efficiency, allowing us to overcome the memory limitations and successfully train our model. 

\subsection{Training}

We opted to train the model using a cloud-rented machine provided by Lambdalabs. The system had 30vCPUs, 200GiB RAM, and 1.4 TiB SSD, and it was equipped with an Nvidia A10 GPU with compute capability of 8.6 and cost about 0.6\$ per hour at the time of writing.

After preprocessing the data, the model was loaded into memory in 8-bit precision and then optimized using Lora with a projection rank of 32. Then the optimized model was put to training for 10 epochs using a batch size of 8 and a learning rate of ${10^{ - 3}}$. 

\section{Results and Discussion}

We used standard evaluation metrics such as accuracy, precision, recall, and specificity. Accuracy is the percentage of the images that were correctly classified. Precision is the accuracy of the positive predictions. The recall is the fraction of the positives that were correctly predicted. F1-Score is the harmonic mean of precision and recall. Specificity is the fraction of the negatives that were correctly predicted.
\begin{center}
    $Accuracy = \frac{TP + TN}{TP + TN + FP + FN}$
\vspace{0.5cm} \\ 
    $Recall = \frac{TP}{TP + FN}$
\vspace{0.5cm} \\ 
    $Precision = \frac{TP}{TP + FP}$
\vspace{0.5cm} \\
    $F1-Score = \frac{2(Precision * Recall)}{Precision + Recall}$
\vspace{0.5cm} \\
    $Specificity = \frac{TN}{TN + FP}$
\end{center}
The results obtained from the binary classification experiment are summarized in table \ref{tab: Results of Binary Classification}. Among the four experiments conducted, the model trained using pathology patients with medium intelligibility levels performed the best, giving an accuracy of 85\%, precision of 0.92, recall of 0.8, F1-score of 0.85, and specificity of 0.91. This indicates that the model finds it easy to predict dysarthria if trained on data containing patients with medium intelligibility levels compared to models trained with other intelligibility levels such as very low, low, and high. Our model achieved around 10\% improvement in accuracy in comparison with the work presented in \cite{ref8}, where they reported a word-level accuracy of 75.3\%.

Table \ref{tab: Output of Multiclass Classification} and Table \ref{tab: Output of Multiclass Digits and Letters Classification} show accuracy, precision, recall, F1-Score, and specificity of models trained on words dataset and digits and letters dataset, respectively for multiclass classification which includes the classes: Control, High, Medium, Low and Very Low. Both models are trained with two patients belonging to each class. The accuracy of multiclass classification using the words dataset is 67\% while the accuracy for multiclass classification using the `digits' and `letters' dataset is 58\%. The multiclass model trained on the words dataset achieved 9\% better accuracy than its counterpart. Analyzing the results from the table, we can see that both models are performing the best in the control class compared with other classes, and both models have a hard time predicting patients from the Low class. Both models have good precision for the high class, but the class has a very low score for other evaluation metrics. 


\begin{table}[!htbp]
  \centering
  \caption{Binary Classification results}
  \label{tab: Results of Binary Classification}
  \begin{tabular}{|p{1.3cm}|p{1cm}|p{1cm}|p{1cm}|p{1.1cm}|p{1.1cm}|}
    \hline
    \textbf{Experiment name} & \textbf{Accuracy (\%)} & \textbf{Precision}  & \textbf{Recall} & \textbf{F1-Score} & \textbf{Specificity} \\
    \hline
    High & 81 & 0.85 & 0.79 & 0.82 & 0.85\\
    \hline
    Medium & 85 & 0.92 & 0.80 & 0.85 & 0.91\\
    \hline
    Low & 76 & 0.78 & 0.77 & 0.77 & 0.77\\
    \hline
    Very-Low & 69 & 0.99 & 0.42 & 0.59 & 0.99\\
    \hline
  \end{tabular}
\end{table}

\begin{table}[!htbp]
  \centering
  \caption{Multiclass Classification Results using words dataset}
  \label{tab: Output of Multiclass Classification}
  \begin{tabular}{|p{1.2cm}|p{1cm}|p{1cm}|p{1.1cm}|p{1.1cm}|}
    \hline
    \textbf{Class} & \textbf{Precision} & \textbf{Recall} & \textbf{F1-Score} & \textbf{Specificity}\\
    \hline
    Control & 0.93 & 0.91 & 0.92 & 0.90\\
    \hline
    High & 0.73 & 0.02 & 0.04 & 0.02\\
    \hline
    Medium & 0.22 & 0.62 & 0.32 & 0.61\\
    \hline
    Low & 0.07 & 0.16 & 0.10 & 0.15\\
    \hline
    Very Low & 0.62 & 0.41 & 0.50 & 0.41\\
    \hline
    \multicolumn{5}{|c|}{Accuracy  =  67\%}\\
    \hline
  \end{tabular}
\end{table}

\begin{table}[!htbp]
  \centering
  \caption{Multiclass Classification Using Digits and Letters dataset}
  \label{tab: Output of Multiclass Digits and Letters Classification}
  \begin{tabular}{|p{1.2cm}|p{1cm}|p{1cm}|p{1.1cm}|p{1.1cm}|}
    \hline
    \textbf{Class} & \textbf{Precision} & \textbf{Recall} & \textbf{F1-Score} & \textbf{Specificity}\\
    \hline
    Control & 0.83 & 0.81 & 0.82 & 0.81\\
    \hline
    High & 1.00 & 0.01 & 0.03 & 0.018\\
    \hline
    Medium & 0.23 & 0.44 & 0.30 & 0.44\\
    \hline
    Low & 0.03 & 0.11 & 0.05 & 0.11\\
    \hline
    Very Low & 0.37 & 0.23 & 0.29 & 0.234\\
    \hline
    \multicolumn{5}{|c|}{Accuracy  =  58\%      } \\
    \hline
  \end{tabular}
\end{table}

\section{Conclusion and Future Works}

This work explores a few-shot learning approach for dysarthria detection using the encoder of the whisper-large-v2 model. The main contributions of the proposed study are:
\begin{itemize}
    \item Compared with the previous methods, pathology detection has improved considerably using transformer models and we are able to demonstrate the potential use of few-shot learning for pathology detection.
    \item From our study we have determined that to detect dysarthria, a model trained using patients having medium-level intelligibility performs better.
    \item We also determined that the dataset built using audio recordings of words will result in better model performance.
\end{itemize}

Potential future works include determining the minimum number of patients to accurately classify dysarthria using few-shot learning and comparative analysis can be done using a wide spectrum of deep learning models to determine which architecture performs the best. 

\bibliographystyle{IEEEtran}
\bibliography{IEEEabrv, refs}
\end{document}